# Frequency conversion cascade by crossing multiple space and time interfaces

Benjamin Apffel[1] and Emmanuel Fort[1]*

[1]Institut Langevin, ESPCI Paris, PSL University, CNRS; 1 rue Jussieu, 75005 Paris, France.
*Corresponding author. Email: emmanuel.fort@espci.fr

**Abstract:**

Time varying media recently emerged as promising candidates to fulfill the dream of controlling the wave frequency without nonlinear effects. However, frequency conversion remains limited by the dynamics of the variations of the propagation properties. Here we propose a new concept of space-time cascade to achieve arbitrary large frequency shifts by iterated elementary transformation cycles. These cycles use an intermediate medium in which wave packets enter and exit through non-commutative space and time interfaces. This concept avoids high frequency or sub-wavelength demanding metamaterials. Upward and downward frequency conversions are performed with 100% efficiency regardless of impedance matching. As an example, we implement this concept with water waves controlled by electrostriction and achieve frequency conversion cascades over a range of 4 octaves.

Spatial control of wave propagation has hardly any limits, while time manipulation remains very challenging. Frequency conversion is an essential component of spectral processing upon which many fundamental researches [1–4] and countless industrial applications [5–7] are based. It is traditionally performed using nonlinear processes which are amplitude dependent and requires high power signals. Time-varying media have recently open exciting perspectives for linear frequency conversion based on wave speed variations [8,9]. This relies on the considerable theoretical developments [10,11] and technological achievements in metamaterials, pervading all types of waves such as electromagnetics [12–19], acoustics [20], elastic [21,22] or hydrodynamics [23,24]. However, large frequency conversions are still challenging as the achievable changes in the medium properties are limited.

Here, we introduce the concept of a "space-time cascade" to perform arbitrary large frequency conversions and circumvent the previous limitations. It consists in iterating arbitrarily-small transformation steps made of a temporal and a spatial interface. The required space-time scaling of the medium variations depends on the space-time extension of the entire wave packet rather than on the wavelength or the frequency of the wave. We implement experimentally this concept with water waves and achieve a frequency conversion cascade over 4 octaves.

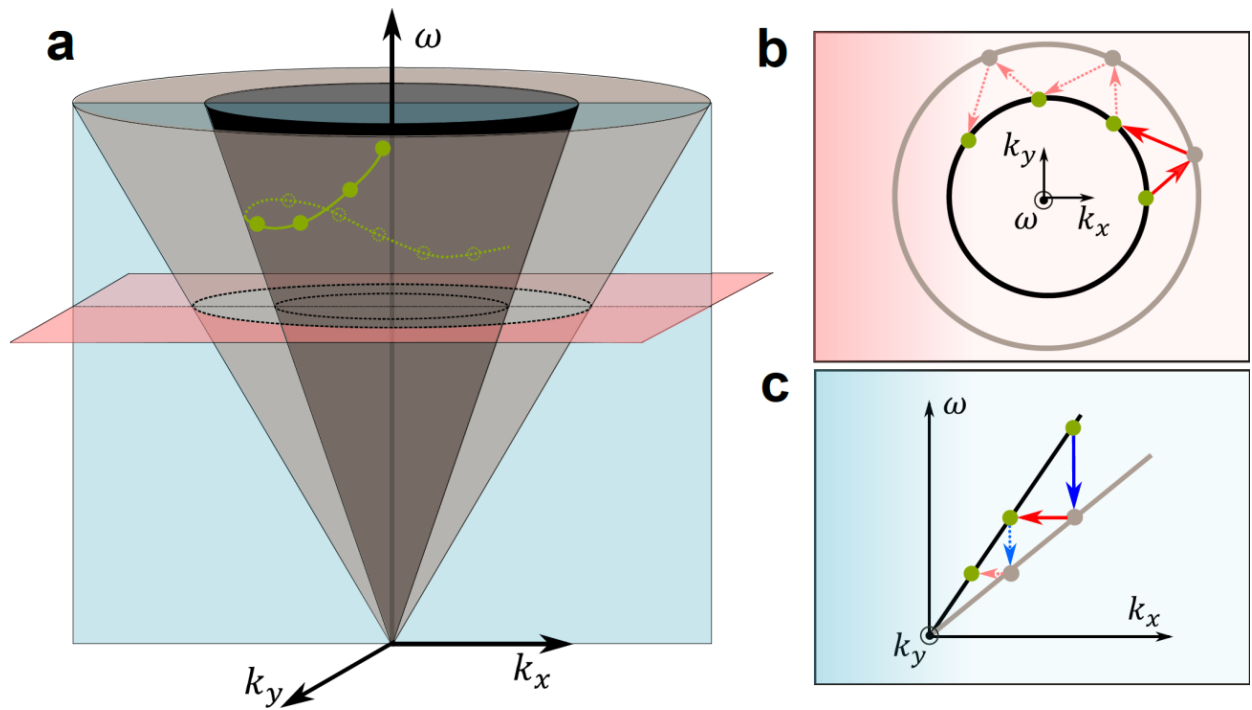

**Figure 1 Principle of wave manipulation by iterative space-time interface crossing. a,** dispersion cones of a reference medium (black) with a ($k, \omega$) wave trajectory representing

wave bending and frequency conversion (green line) achieved by successive elementary transformations through space and time interface crossing with an intermediary medium with a slightly different dispersion cone (grey). **b**, A plane wave $(\boldsymbol{k_1}, \omega_1)$ in medium #1 is projected to $(\boldsymbol{k_2}, \omega_2)$ when crossing an interface between medium #1 and #2. For a plane space interface, the projection is horizontal along its normal direction $\boldsymbol{u} \propto \boldsymbol{k_2} - \boldsymbol{k_1}$ and $\omega_2 = \omega_1$ (red arrows). Elementary horizontal transformation (solid line arrows) composed of two space interface crossings with different directions can be iterated to achieve arbitrary wave bending (dashed arrows). **c**, For a time interface, the projection is vertical with $\boldsymbol{k_2} = \boldsymbol{k_1}$ and $\omega_2 \neq \omega_1$ (blue arrows). The iteration of an elementary vertical frequency conversion transformation composed of a time interface projection followed by a space interface projection (solid arrows) generates arbitrary frequency conversions (dashed arrows). The reflected waves have been omitted for clarity (see Supplementary Information).

The idea of space-time cascade is conceptually sketched in Fig. 1a. A wave packet centered around the angular frequency $\omega_1$ can be represented to a good approximation as a point $(\boldsymbol{k_1}, \omega_1)$ on the dispersion cone of medium #1, with $\boldsymbol{k_1}$ being its associated wave vector (Fig. 1a). Comprehensive wave control involves the ability to change the direction and frequency of a wave, i.e., to move the characteristics of the wave from one point $(\boldsymbol{k_1}, \omega_1)$ to another arbitrary point $(\boldsymbol{k'_1}, \omega'_1)$ on the dispersion cone. Such a shift can be achieved by using an intermediate medium #2. As the wave packet moves from medium #1 into #2, $(\boldsymbol{k_1}, \omega_1)$ is projected at $(\boldsymbol{k_2}, \omega_2)$ on the dispersion cone of medium #2. The projection transformation depends on the characteristics of the interface separating the two media through which the wave packet passes. For instance, a planar space interface yields a horizontal projection, $S_{1 \to 2}$, in the direction normal to the interface (Fig. 1b). Spatial and time translation invariance results in the conservation of the wave vector component tangent to the interface and $\omega_2 = \omega_1$, respectively. Projections associated with time interfaces, $T_{1 \to 2}$, are obtained by time varying the propagation properties from medium #1 to #2 [8,17,19]. In this case, momentum is conserved ($\boldsymbol{k_2} = \boldsymbol{k_1}$) while the frequency changes $\omega_2 \neq \omega_1$ (Fig. 1c). The wave packet transmitted in medium #2 can be projected back to medium #1 but through a different interface. Since the projections depend on the interfaces, the transformation cycle is not reduced to the identity. For example, passing through two spatial interfaces of different directions induces wave bending with a change in direction $\boldsymbol{\Delta k}$ at fixed $\omega$ (Fig. 1b). Transformations inducing frequency conversion require a combination of a time interface and a space interface such as $T_{2\to1}S_{1\to2}$ or $S_{2\to1}T_{1\to2}$. Pure frequency conversion transformation are obtained for vertical step between the dispersion cones when the space interface is oriented normal to the initial wave vector forms a (Fig. 1c).

In practice, the dispersion cones of the two media are very close to each other due to the difficulty of significantly varying the propagation properties of a medium. Step transformations therefore induce very limited frequency shifts $\Delta \nu$. However, these small shifts can be added by iteration (Fig. 1c) allowing arbitrarily large transformations between two media with arbitrarily close properties. Any trajectory on the dispersion cone characterizing wave bending and frequency conversion can be implemented using this concept (Fig. 1a). Frequency conversions can be achieved by cascading elementary vertical steps to form a staircase-like transformation with a priori no limitation. A simple permutation of space and time interfaces can change a blue-shift into a redshift, going up or down the transformation staircase. The commutated product, changing $T_{2\to1}S_{1\to2}$ to $S_{2\to1}T_{1\to2}$, corresponds to the time-reversed transformation. Since the transformations are linear, the previous considerations can easily be extended to wave packets with arbitrary frequency spectra.

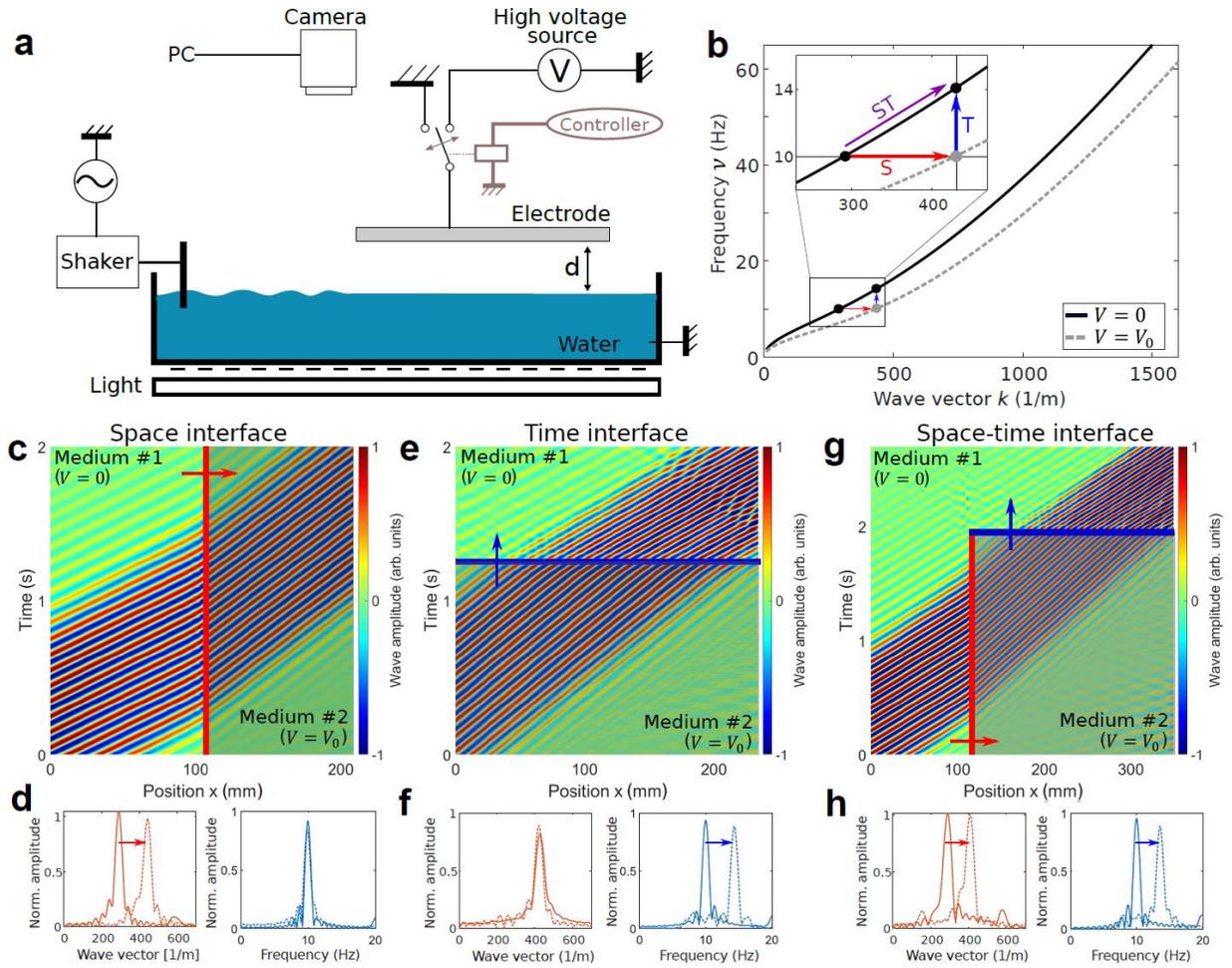

**Figure 2 Experimental implementation of elementary space-time transformation step. b**, Schematic of the experimental setup. The shaker produces a plane water wave packet. The transparent electrode changes medium #1 into #2 by electrostriction with controllable high voltage switches. The wave field is measured by a camera from the deformation of a checkerboard pattern placed under the container (see Supplementary Information). **b**, Water waves dispersion curves of media #1 with no voltage and #2 with $V_0 = 7.5$ kV. Example of an elementary frequency up-conversion transformation step composed of successive space projection $S_{1\to 2}$ and time projection $T_{2\to 1}$ for an initial wave at $\nu = 10$ Hz. Normalized kymographs of a wave packet centered at $\nu = 10$ Hz crossing **c**, a space interface $S_{1\to 2}$, **e**, a time interface $T_{2\to 1}$ and **g**, the elementary transformation step $T_{2\to 1}S_{1\to 2}$ (see Supplementary Video 1 and Information). **d**, **f** and **h** normalized $k$-spectra and $\nu$-spectra taken before and after the interface crossings, measured from (c), (e) and (g) respectively.

We implemented this concept with electrostriction-controlled water waves. When an flat electrode is mounted above the grounded conductive water surface, the applied electric field exerts an attractive force on the liquid surface that changes the velocity of the water wave [25]. This creates well-controlled space and time varying properties. Figure 2a shows the experimental setup consisting of a container filled with tap water (see Supplementary Information). Transparent ITO electrodes can be suspended horizontally at a distance $d$ above the water. The electric potential $V$ can be tuned in the range of 0 to 10 kV using electrical switches. Plane waves are produced by a shaker exciting horizontally a paddle. The wave field is measured from top-view images using the deformation of a checkerboard pattern placed below [26]. For a given wavenumber $k$, the voltage-dependent refractive index $n(k,V)$ satisfies

$$n(k,V) = (1 - \chi_0(k)V^2)^{-1/2} \qquad \text{with } \chi_0(k) = \epsilon/(\rho c^2(k)d^2 \tanh(kd)) \qquad (1)$$

$c(k)$ is the wave velocity given by the gravity-capillary dispersion relation at $V = 0$, $\epsilon$ is the dielectric permittivity of air and $\rho$ is the density of the liquid [25]. Figure 2b shows the dispersion relation (1) for $V = 0$ kV (medium #1) and for $V = V_0$ (medium #2) as well as the frequency up-conversion transformation step consisting of a space interface $S_{1 \to 2}$ followed by a time interface $T_{2 \to 1}$. The space interface is located at the edge of the electrode where the refractive index varies typically over a width $\sim d$. Figure 2c shows the experimental normalized kymograph of a wave packet crossing the interface $S_{1 \to 2}$ as it enters under the electrode set at $V_0$ to produce a change of refractive index $\Delta n \approx 0.5$ (Supplementary Video 1). The wave vector is shifted by $\Delta k \approx +200$ m$^{-1}$ while its frequency spectrum remains unchanged (Fig. 2d). The wave packet propagating under an electrode can also cross a time interface $T_{2 \to 1}$ when the voltage $V_0$ is switched off, resulting in a sudden change $\Delta n \approx -0.5$ (Supplementary Video 1). The frequency spectrum is blue-shifted by $\Delta \nu \approx +4$ Hz while the $k$-spectrum remains unchanged (Fig. 2f). The complete elementary transformation step consisting of a succession of interfaces $S_{1 \to 2}$ and $T_{2 \to 1}$ (Fig. 2g) shifts both the frequency and the wave vector (Fig. 2h) to satisfy the dispersion relation of medium #1. The kymographs show transformation steps as space-time tessellation. Medium #2 appears as square patterns which size depends on the space-time extension of the wave packet (Supplementary Fig. 1).

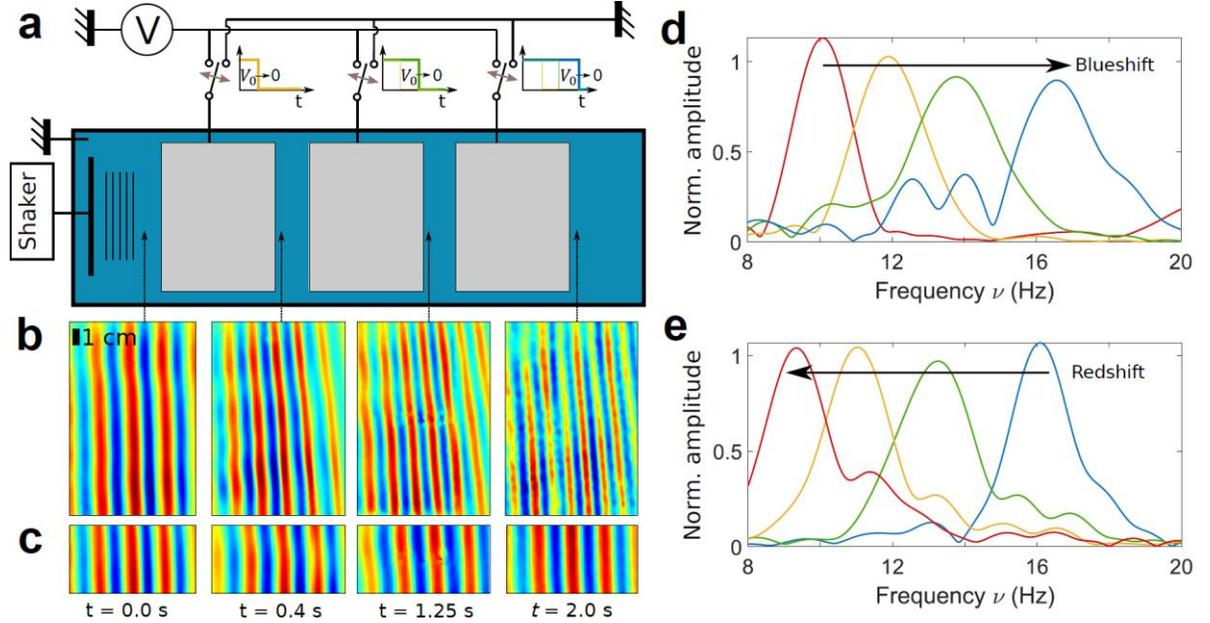

**Figure 3 Experiment of the frequency conversion cascade. a**, Schematic of the space-time cascade experimental setup (top view) with 3 electrodes controlled by independent electrical switches. The plane wave packet with $\nu \approx 10$ Hz undergoes 3 successive elementary transformation steps $T_{2\to1}S_{1\to2}$ as it passes under each electrode initially set at $V_0$ and switched off synchronized with the wave propagation (see insets). **b**, Snapshots of the initial wave packet ($t = 0.0$ s) and of the successive frequency shifts after each transformation step at $t = 0.4$ s, 1.25 s and 2.0 s (Supplementary Video 2). **c**, Snapshots of the same propagating wave packet taken in the absence of voltage (reference). **d**, Frequency spectra of the snapshots (B) showing a blue-shift $\Delta\nu \approx 2$ Hz for each transformation step. **e**, Frequency spectra associated with the time reversed process for an initial wave packet of $\nu \approx 16$ Hz redshifted by successive elementary transformation steps $S_{2\to1}T_{1\to2}$ (Supplementary Video 3 and Fig. 1).

The experimental setup can be modified to perform frequency conversion cascades (Fig. 3a). The wave packet undergoes multiple transformation steps $T_{2\to1}S_{1\to2}$ as it propagates under successive electrodes. The wavelength $\lambda$ of the wave packet decreases of after each step (Fig. 3b compared to a free propagating reference, Fig. 3c). These contractions are associated blue-shift in frequency satisfying $\nu = \frac{c}{\lambda}$ (Fig. 2b and Supplementary Fig. 1 and Video 2). The frequency spectrum is shifted by $\Delta\nu \approx 2$ Hz at each step (Fig. 3d). The opposite redshift cascade can also be achieved by the time reversed operation with successive time-flipped steps $S_{2\to1}T_{1\to2}$ (Fig. 3e, Supplementary Video 3).

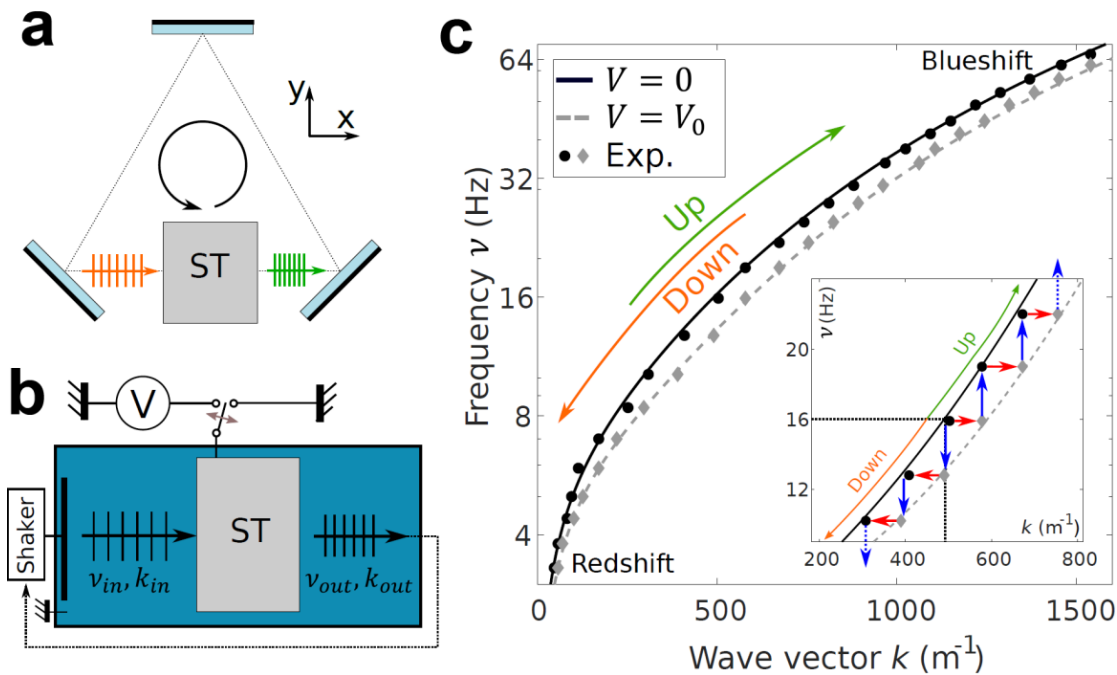

**Figure 4 Frequency conversion cascade confined within a cavity. a**, Example of a cavity with a triangle geometry to perform multiple passes under the same electrode. The circulating wave packet is limited to the size of active medium (ST) with no links to the cavity modes. **b**, Experimental implementation in a modified Fabry-Pérot cavity with an amplification loop to compensate for the poor reflection efficiency of water waves. **c**, Up and down frequency conversion cascades and dispersion curves from eq. (1) in semi-logarithmic scale for an initial wave packet at $\nu = 16$ Hz. (inset: close-up).

Other geometries are more suitable when the number of transformation steps increases. Cavities, such as a unidirectional triangular geometry (Fig. 4a) or a Fabry-Pérot cavity, allow a single active medium to perform all the transformation steps. The time interface is synchronized when the circulating wave packet is in the active medium. Since the extension of the wave packet is limited, it is completely independent of the cavity modes in contrast to time-varying resonators experiments [9,12,13,27]. Due to the limited efficiency of water waves reflection an amplification loop must be used to send the wave packet back into the active medium (Fig. 4b). With damping compensation, a cascade of up and down frequency conversion can be achieved over a range of more than 4 octaves with 23 elementary steps (Fig. 4c).

As in most time-varying experiments, electrostriction-driven interfaces have an impedance mismatch creating reflected waves. These small amplitude waves can be further reduced by anti-reflection coating [17] or by using smoother interfaces [28]. In the space-time cascade, the duration of time interfaces is limited only by the residence time of the wave packet in the active medium. This is an asset for implementation with other types of waves such as in optics.

What is the influence of the impedance mismatch on the energy conversion yield of a transformation step? Energy being conserved in space interfaces, the impedance mismatch is detrimental reflecting part of the energy. For time interfaces, momentum is conserved, but not energy. Hence, the amplitude of the transmitted wave increases with impedance mismatch to balance the reflected momentum. For sharp space and time interfaces, the two effects compensate and the energy flux density yield for an elementary transformations step satisfies $\Gamma = \omega'/\omega$, independent of the order of the projections (Supplementary Information)[8]. Whatever the impedance mismatch, the yield for sharp interfaces is equal to that of a perfectly matched interface for which the entire wave packet is transmitted with 100% efficiency. From a photonic viewpoint, the energy of each photon is shifted from $\hbar\omega$ to $\hbar\omega'$ but the total number of transmitted photons remains constant. A different shaping of the space and time interfaces could even lead to an amplification of the output wave packet.

Along with the transformation optics [29,30], the space-time cascade also find an interesting interpretation in terms of the space-time metric of general relativity. In the ray approximation, the light beam is characterized by a 4-momentum vector $p^\mu \propto \left(\frac{\omega}{c}, -\boldsymbol{k}\right)$ tangent to the geodesic along which it propagates. Observers placed along the geodesic would measure in their rest frame changes in directions and frequencies following a $(\boldsymbol{k}, \omega)$ curve on their common dispersion cone $\omega = ck$ (Fig. 1a). A space-time cascade can thus mimic large light–bending and frequency conversions associated to gravitational effects or universe expansion.

4. Milton, G. W. & Mattei, O. Field patterns: a new mathematical object. *Proc. R. Soc. A.* **473**, 20160819 (2017).

5. Matsuda, N. Deterministic reshaping of single-photon spectra using cross-phase modulation. *Science Advances* **2**, e1501223 (2016).

6. Ramaccia, D., Alù, A., Toscano, A. & Bilotti, F. Temporal multilayer structures for designing higher-order transfer functions using time-varying metamaterials. *Appl. Phys. Lett.* **118**, 101901 (2021).

7. Nassar, H. *et al.* Nonreciprocity in acoustic and elastic materials. *Nature Reviews Materials* **5**, 667–685 (2020).

8. Morgenthaler, F. Velocity Modulation of Electromagnetic Waves. *IRE Transactions on Microwave Theory and Techniques* **6**, 167–172 (1958).

9. Notomi, M. & Mitsugi, S. Wavelength conversion via dynamic refractive index tuning of a cavity. *Phys. Rev. A* **73**, 051803 (2006).

10. Xiao, Y., Agrawal, G. P. & Maywar, D. N. Spectral and temporal changes of optical pulses propagating through time-varying linear media. *Opt. Lett.* **36**, 505–507 (2011).

11. Huidobro, P. A., Galiffi, E., Guenneau, S., Craster, R. V. & Pendry, J. B. Fresnel drag in space–time-modulated metamaterials. *Proc Natl Acad Sci USA* **116**, 24943–24948 (2019).

12. Tanabe, T., Notomi, M., Taniyama, H. & Kuramochi, E. Dynamic Release of Trapped Light from an Ultrahigh-Q Nanocavity via Adiabatic Frequency Tuning. *Phys. Rev. Lett.* **102**, 043907 (2009).

13. Preble, S., Xu, Q. & Lipson, M. Changing the Color of Light in a Silicon Resonator. *Nature Photonics* **1**, 293–296 (2007).

14. Shaltout, A. M. *et al.* Spatiotemporal light control with frequency-gradient metasurfaces. *Science* **365**, 374–377 (2019).

15. Bruno, V. *et al.* Broad Frequency Shift of Parametric Processes in Epsilon-Near-Zero Time-Varying Media. *Applied Sciences* **10**, 1318 (2020).

16. Zhou, Y. *et al.* Broadband frequency translation through time refraction in an epsilon-near-zero material. *Nature Communications* **11**, 2180 (2020).

17. Pacheco-Peña, V. & Engheta, N. Antireflection temporal coatings. *Optica* **7**, 323–331 (2020).

18. Silva, A. *et al.* Performing Mathematical Operations with Metamaterials. *Science* **343**, 5 (2014).

19. Caloz, C. & Deck-Léger, Z.-L. Spacetime Metamaterials—Part I: General Concepts. *IEEE Transactions on Antennas and Propagation* **68**, 1569–1582 (2020).
10

**Acknowledgments:** The authors would like to thank Mathias Fink and Antonin Eddi for fruitful discussions and Samuel Hidalgo for his precious help in setting the electric part of the experimental setup. The authors thank the support of AXA research fund and the French National Research Agency LABEX WIFI (ANR-10-LABX-24).

**Competing interests:** Authors declare no competing interests.

**Additional information:** Supplementary Information and Videos are available for this paper. Correspondence and requests for materials should be addressed to E. F.

# Supplementary Information

**Supplementary Materials and Methods:**

1. **Experimental setup**

Experiments are performed in a Plexiglas tank of $100 \times 50 \times 20$ cm. The water depth is always greater than 7 cm so that the deep water approximation is always valid in our range of frequencies.

The transparent electrodes are made of glass plates coated with a thin ITO layer. The resistance between two points of the electrode is always measured to be less than $100 \, \Omega$ ensuring that potential applied is the same on the whole electrode.

The high-voltage power supply (Spellman MPS15P10/24) delivers a constant voltage typically between 8 kV and 10 kV. The electrodes are switched from high-voltage to ground using high-voltage relays (Meder HM24) controlled with an Arduino Controllino Mini.

The tilt of the electrode with respect to the water surface is set with three micrometric screws. The horizontality of the electrodes is controlled by sending a collimated laser beam on the electrode and the water surface with a direction close to the normal of the two interfaces. The two reflected beams then pass through a converging lens with focal $f \approx 30$ cm and are imaged on a screen placed in the image focal plane. When the two focal spots are at the same position, the angle between the electrode and the surface was always measured to be less than 0.1°.

Dish soap (Paic Citron) is used to lower the surface tension and change the dispersion relation. This amplifies the effect of electric potential on the change of the wave speed, ie on the change of refractive index $n$. The latter being the relevant parameter, the exact value of the surface tension is not needed but can be deduced from dispersion curve fitting.

2. **Measurement of the wave amplitude**

Water waves are filmed from above at 100 fps using a Basler camera. A checkerboard pattern is placed under the water tank in order to reconstruct numerically the waves using FCD Schlieren method *(26)*. For each time, we recover the amplitude $A(x, y, t)$ on the area of interest. A 2D high-pass spatial filter is applied to each image $A(x, y)$ and a 1D high-pass temporal filter is applied in the time domain. The waves are plane waves propagating along the $x$ axis (horizontal axis on the images and movies). The spatio − temporal diagrams $A(x, t)$ is obtained by averaging the wave amplitude along the wave front $y$-direction (vertical in the images and movies).

In order to compensate for damping along the propagation, the amplitude $A(x, t)$ at each $x$ is divided by the standard deviation $\sigma(x) = \sqrt{\sum_t A(x,t)^2}$ taken at the same $x$ in the time direction.

By taking slices along the $x$ (resp. $t$) direction and performing Fast Fourier Transform, it is possible to determine the wave vector (resp. the frequency) at a given time (resp. position). To increase the precision on the determination of $\omega$ or $k$ corresponding to the maximum of the Fourier transform, zero-padding is used so that each sample contains 2048 points.

This procedure was applied to obtain the data presented Fig. 2 in the main text. The spectra of Fig. 3 in the main text were extracted from the diagram presented Fig. S1.

For the case of temporal interfaces, changing the voltage of the electrodes generates water waves on the bath. These water waves are removed by subtracting a movie containing only the waves emitted by the electrodes.

### 3. Dispersion relation for water waves under an electrode

We consider a wave of wave vector $k$ and pulsation $\omega$ at the surface of water. The gravito-capillary dispersion relation is $\omega_0^2 = gk + \gamma k^3/\rho$ with $g = 9.8$ m/s² the local gravity, $\gamma \approx 20$ mJ/m² the surface tension and $\rho = 10^3$ kg/m3 the density. When a horizontal electrode is placed at distance $d$ and potential $V_0$ over the grounded water, the dispersion relation is modified as (25)

$$\omega_1^2 = \omega_0^2(k)\left(1 - \frac{\epsilon V_0^2 k^2}{d^2 \rho \omega_0^2(k) \tanh(kd)}\right) \tag{S1}$$

with $\epsilon = 9.10^{-12}$F/m the dielectric constant of air. The previous relation has been verified experimentally in the frequency range used for our frequency cascade. We cover partially the surface with a horizontal electrode at potential $V_0$. A wave at frequency $\nu = 10, 14$ or $18$ Hz is sent on a bath. When the waves goes from the free surface to the area covered by the electrode, the wave vector changes from $k_0$ to $k_1$ while the frequency $\nu$ remains the same. For each frequency, the ratio $n_{\text{exp}} = k_1/k_0$ are measured experimentally for different values of $V_0$. The results are shown in Fig. S2.

The theoretical ratio $n_{\text{th}} = k_1/k_0$ given by eq. for $\nu = 10, 14$ and $18$ Hz is plotted Fig. 1 with $\gamma = 23$ mJ/m² and $d = 6.7$ mm. The surface tension $\gamma$ has been obtained by measuring $k_0$ for $\nu = 18$ Hz. The distance $d$ was measured experimentally to be $d \approx 6$ mm and was adjusted to fit the datas. We see that for our frequency range, the theory provides satisfying prediction with absolute difference $|n_{\text{th}} - n_{\text{exp}}| < 0.1$. Note that the value $\gamma = 23$mJ/m² differs from the one used in the main text ($\gamma = 45$mJ/m²) because experimental condition (in particular the amount of soap used) were different in both cases. This variation is however not an issue as only the effective value of $n$ is significant for our experiments.

### 4. Reflection/transmission amplitude coefficients for a sharp space and time interfaces

The Fresnel coefficients associated to the amplitude of a monochromatic electromagnetic wave field can be obtained writing continuity equation across the interface between media #1 and #2 with impedances $\eta_i$ and permittivity $\epsilon_i$ with $i = 1$ and 2.
In the case of a sharp interface the equations are (see ref. (8)):
For a space interface, the transmission and the reflection coefficients are $t_{12}^S = \frac{2\eta_2}{\eta_1+\eta_2}$ and $r_{12}^S = \frac{\eta_2-\eta_1}{\eta_1+\eta_2}$ respectively.

For a time interface, the transmission and the reflection coefficients are $t_{12}^T = \frac{\epsilon_1}{\epsilon_2}\frac{\eta_1+\eta_2}{2\eta_2}$ and $r_{12}^T = \frac{\epsilon_1}{\epsilon_2}\frac{\eta_2-\eta_1}{2\eta_2}$ respectively.

### 5. Transmission yield for an elementary transformation step

An elementary transformation step consists in a succession of a time and a space interfaces which order results in an upward or downward frequency shift.
We consider a quasi-monochromatic wave packet with amplitude $A$ at $(k, \omega)$ changed into another wave packet with amplitude $A'$ at $(k', \omega')$ after an elementary transformation using medium #2 as an intermediary medium.
We neglect the dispersion in medium #1 so that the impedance refractive index $n_1$ is the same for $\omega$ and $\omega'$. Since the frequency shift of an elementary transformation step is small this is usually satisfied.

In the case of an elementary step with first a space interface followed by a time interface, the change in amplitude satisfies

$$\frac{A'}{A} = t_{21}^T t_{12}^S = \frac{n_1}{n_2}$$

Since $\frac{c}{n_1} = \frac{\omega}{k} = \frac{\omega'}{k'}$ and $\frac{c}{n_2} = \frac{\omega}{k'}$, it gives $\frac{A'}{A} = \frac{k'}{k} = \frac{\omega'}{\omega}$

In the case of an elementary step with first a time interface followed by a space interface, the change in amplitude satisfies

$\frac{A'}{A} = t_{21}^S t_{12}^T = \frac{n_2}{n_1}$

Since $\frac{c}{n_1} = \frac{\omega}{k} = \frac{\omega'}{k'}$ and $\frac{c}{n_2} = \frac{\omega'}{k}$, it gives $\frac{A'}{A} = \frac{k'}{k} = \frac{\omega'}{\omega}$

Hence, the change of amplitude of the wave packet is independent of the order of the projections and of the impedances of the two media.

The wave packet changes its size as it undergoes a transformation steps due to the change of refractive index when crossing the space interface. It is associated to the change in wavelength and satisfies $\frac{\lambda'}{\lambda} = \frac{k}{k'}$.

It follows that the yield of the energy density flux $\Gamma$ for the transformation step is given by

$\Gamma = \frac{\lambda' A'^2}{\lambda A^2} = \frac{k'}{k} = \frac{\omega'}{\omega}.$

The energy density flux is again independent of the impedance of the two media and of the order of the projections. Note that $\Gamma > 1$ if $\omega' > \omega$ i.e. in case of a blueshift which is not in contradiction with energy conservation since energy can be produced at the time interface.
(see figure S4).

## 6. Experimental observation of transmitted and reflected wave packets

We evaluate experimentally these Fresnel coefficients. We first produce a wave packet that propagates on the free surface without crossing any interface. We then produce the same wave packet and measure the wave field as it crosses an interface. By comparing the two wave packets, we can deduce the reflection and transmission coefficients for a space or a time interface. Experimental results are shown Fig. S3 for the spatial interface and Fig. S4 for the temporal interface with $\Delta n = 0.5$. This corresponds to the largest refractive index variation achievable with our experiment. In both cases, we see that the wave amplitude is only marginally changed by the interface (Fig. S3b and S4b). This suggests that that the transmission coefficient is $t_{12}^T \approx t_{12}^S \approx 1$ while the reflexion coefficient is very close to zero $r_{12}^T \approx r_{12}^S \approx 0$. This is confirmed when one looks at the space-time diagrams (Fig. S3c and S4c). The reflection coefficient would be associated to waves going in the opposite direction compared to the transmitted one (Fig. S3a and S4a). However, these waves are not visible on the diagram. If one take the Fourier transform of the diagram, the peak corresponding to the reflected wave is at least one order of magnitude smaller than the peak corresponding to the transmitted field. Similar observations have been made for lower values of $n$.

The Fresnel coefficients associated with a variation of permittivity only would give $t_{12}^S = 0.8, r_{12}^S = -0.2, t_{21}^T = 1.85$ and $r_{21}^T = 0.375$. From the experimental observations, it is clear that the water wave interfaces produced by electrostriction have a much better impedance matching. The transmission/reflection coefficients also depend on the exact space and time profile of the interface. In our case, it is difficult to find analytical expressions for the wave at the interface. This would require the potential at the boundary of the electrode while the water surface height changes in space. As the distance between the electrode and the water surface is of the order of the wavelength and the capillary length, the problem cannot be simplified easily. For the time interface, the water response to an abrupt change of the electric potential is also a complex problem. However, since the yield of an elementary step is independent of the impedance mismatch, the detail of the time and space interfaces is not essential. We nevertheless show experimentally that in this configuration, the reflected waves can be ignored with a very good approximation and the transmitted wave amplitude is close to unity.

**Supplementary Figures:**

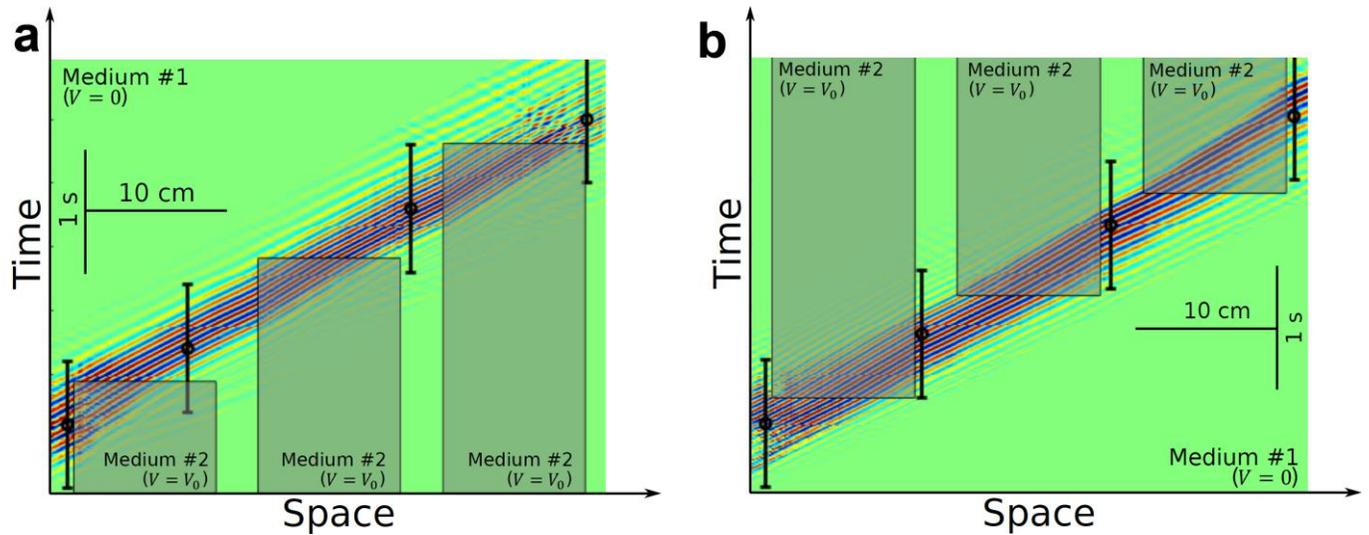

**Figure S1 Upward and downward frequency conversion cascades.** Normalized kymographs of a wave packet centered at $\nu = 16$ Hz crossing 3 successive elementary transformation steps (**a**) for a blueshift with steps $T_{2\to1}S_{1\to2}$ and (**b**) for a redshift with steps $S_{2\to1}T_{1\to2}$. These kymographs are associated to Fig. 2b-2d in the main text and Video S2 for (a) and Fig. 2e in the main text and Video S3 for (b).

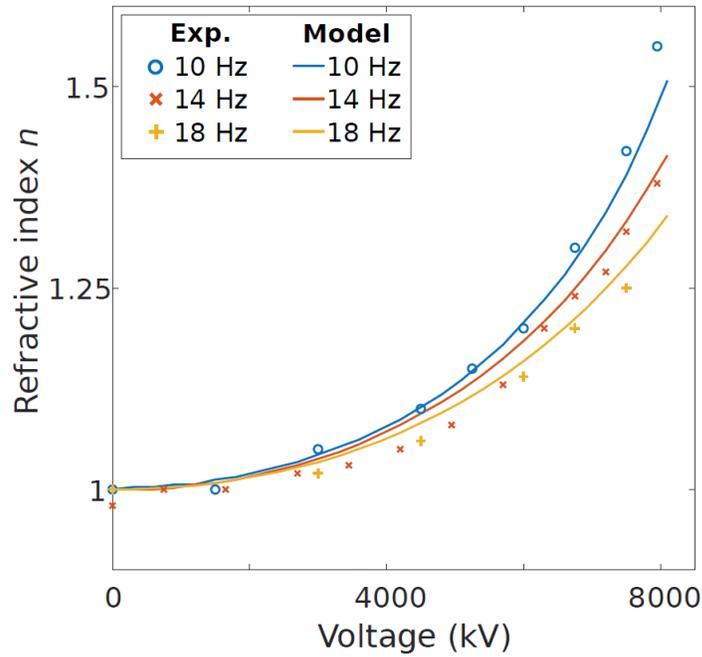

**Figure S2 Refractive index controlled by electrostriction.** Experimental measurements of the index $n$ for wave frequency of 10 Hz (blue circles), 14 Hz (red x-crosses) and 18 Hz (yellow +-crosses) as a function of the applied voltage under the electrode. The associated curves are theoretical predictions from the modified dispersion relation eq. (S1) with $d = 6.7$ mm and $\gamma = 23$ mJ/m².

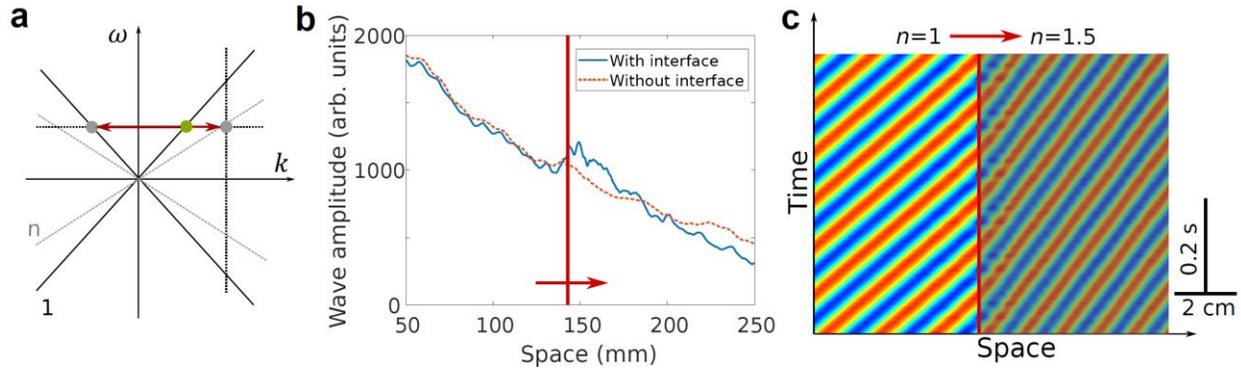

**Figure S3 Reflection and transmission on the space interface. a**, Schematic of the crossing of a space interface in the wave number/angular frequency plane. A plane wave represented as a point $(k_1, \omega_1)$ (green circle) on the dispersion curve of medium #1 with refractive index 1 (black solid line) produces a transmitted wave in medium #2 with refractive index $n$ (grey circle on the grey dotted line) and a reflected wave in medium #1 (grey circle on the black solid line). The former is at position $(k_2 = nk_1, \omega_2 = \omega_1)$, the latter is at position $(-k_1, \omega_1)$. **b**, Experimental measurement of the wave amplitude propagating across a space interface (solid blue line) as compared with the reference amplitude for the same wave with no space interface (red dashed line). The amplitude of the wave at each position is measured by taking the standard deviation in time. **c**, Experimental space-time diagram of the monochromatic wave at $\nu = 10$ Hz crossing a space interface with $n = 1.5$. The reflected wave is not detected and the

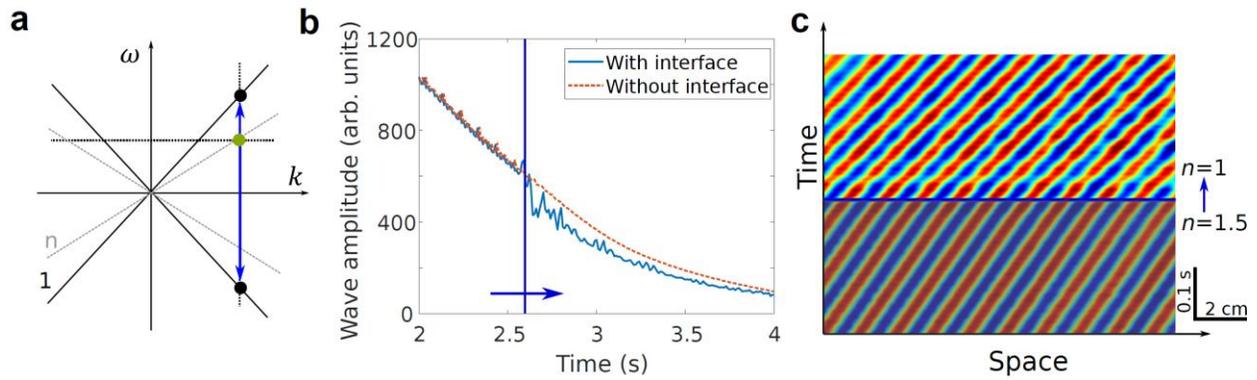

**Figure S1 Reflection and transmission on the time interface. a**, Schematic of the crossing of a time interface in the wave number/angular frequency plane. A plane wave represented as a point $(k_1, \omega_1)$ (green circle) on the dispersion curve of medium #2 with refractive index $n = 1.5$ (grey dotted line) produces a transmitted wave (upper grey circle) and a reflected wave (lower grey circle) in medium #1 with refractive index 1. The former is at position $(k_2 = k_1, \omega_2 = n\omega_1)$, the latter is at position $(k_2 = k_1, \omega_2 = -n\omega_1)$. **b**, Experimental measurement of the wave amplitude propagating across a time interface (solid blue line) as compared with the reference amplitude for the same wave with no time interface (red dashed line). The amplitude of the wave at each time is measured by taking the standard deviation in space. **c**, Experimental space-time diagram of an initial monochromatic wave at $\nu = 10$ Hz crossing a time interface with $\Delta n = -0.5$.